\documentclass{aa}
\usepackage[varg]{txfonts}

\usepackage{graphicx}
\usepackage{dcolumn}
\usepackage{bm}

\usepackage{amsmath}
\usepackage{amssymb}

\usepackage{natbib}

\newcommand{\ep}{\epsilon}

\newcommand{\cvir}{c_{vir}}
\newcommand{\rvir}{R_{vir}}

\renewcommand{\theenumi}{\Roman{enumi}}

\begin{document}

\titlerunning{Is the Navarro-Frenk-White profile real?}

\title{The real and apparent convergence of N-body simulations of the dark matter structures: is the Navarro-Frenk-White profile real?}

\author{A. N. Baushev}
\offprints{baushev@gmail.com}
\author{Anton N. Baushev\inst{1, 2}}
\institute{DESY, 15738 Zeuthen, Germany\and
 Institut f\"ur Physik und Astronomie, Universit\"at Potsdam, 14476
Potsdam-Golm, Germany}

\date{}

\abstract{}


\abstract{While N-body simulations suggest a cuspy profile in the centra of the dark matter halos
of galaxies, the majority of astronomical observations favour a relatively soft cored density
distribution of these regions. The routine method of testing the convergence of N-body simulations
(in particular, the negligibility of two-body scattering effect) is to find the conditions under
which the shape of the formed structures is insensitive to numerical parameters. The results
obtained with this approach suggest a surprisingly minor role of the particle collisions: the
central density profile remains untouched and close to the Navarro-Frenk-White shape, even if the
simulation time significantly exceeds the collisional relaxation time $\tau_r$. In order to check
the influence of the unphysical test body collisions we use the Fokker-Planck equation. It turns
out that a profile $\rho\propto r^{-\beta}$ where $\beta\simeq 1$ is an attractor: the
Fokker-Planck diffusion transforms any reasonable initial distribution into it in a time shorter
than $\tau_r$, and then the cuspy profile should survive much longer than $\tau_r$, since the
Fokker-Planck diffusion is self-compensated if $\beta\simeq 1$. Thus the purely numerical effect of
test body scattering may create a stable NFW-like pseudosolution. Moreover, its stability may be
mistaken for the simulation convergence. We present analytical estimations for this potential bias
effect and call for numerical tests. For that purpose, we suggest a simple test that can be
performed as the simulation progresses and would indicate the magnitude of the collisional
influence and the veracity of the simulation results.}


\maketitle

\section{Introduction}
\subsection{The 'cusp vs.\ core' problem}
N-body simulations is one of the most prominent methods of investigation of the universe structure
formation: the complexness of the task makes any detailed analytical consideration almost hopeless.
The rapid growth of computational capability, as well as the algorithm improvement, have made the
progress in this field really impressive. Despite of it, there are still difficulties to be solved:
the problem of the density profile in the central region of formed dark matter (DM) haloes is
probably the most important and intriguing among them. N-body simulations always suggest a very
steep, cuspy profile in the center (Navarro-Frenk-White (hereafter NFW) or Einasto with high power
index \citep{moore1999, neto2007, mo09, navarro2010}). The NFW profile, for instance, has an
infinite density in the center
\begin{equation}
 \rho=\dfrac{\rho_s r^3_s}{r(r+r_s)^2}
 \label{13a1}
\end{equation}
Meanwhile, observations strongly favor a cored profile, i.e. a respectively shallow density
distribution at $r\to 0$ \citep{deblok2001, bosma2002, marchesini2002, gentile2007, 11, mamon2011}.

The cusp disappearance might be accounted for by the influence of the baryon component. However,
observations suggest that the cusp is absent in dwarf satellites of the Local Group galaxies
 having only a very minor fraction of baryons \citep{oh2011, governato2012, tollerud2012}. The
conflict between the simulations and the observations calls for additional verification of the
simulation accuracy and convergence.

\subsection{N-body simulations and relaxation processes}
The idea of N-body simulations is to substitute real dark matter particles by heavy test bodies, so
that the average density remains the same. It allows to decrease the number of particles and make
the task calculable. In order to avoid unphysical close encounters, the Newtonian potential of the
bodies is cut on short distances: potential is set to be constant inside some radius $\varrho$. The
initial conditions of the cosmological simulations are usually chosen as a random Gaussian field,
so that they model the real initial cosmological perturbations. Then the system freely evolves
forming structures, including haloes of various masses. It is important to mention that, though the
total number of test particles exceeds $\sim 10^9$ at present, a single formed halo rarely contains
more than $\sim 10^6$ simulation particles in this type of simulations.

It was shown in \citep{15, 16} that the energy evolution of the system plays the key role in the
cusp formation: a core inevitably occurs in the center, if the energy relaxation of the system is
moderate; a significant part of the halo particles should change their energies many ($\sim
c_{vir}$) times with respect to the initial values, in order to form a cusp. This requirement is
rather weak in the case of galaxy clusters with low NFW concentrations ($c_{vir}\simeq 3-6$), but
implies the energy change by at least an order of magnitude in the case of galactic haloes
($c_{vir}\simeq 12-20$).

A dark matter halo has several ways of relaxation. First of all, the baryon matter can affect this
process, though dwarf galaxies have only a tiny fraction of the baryons and show no cusps
\citep{oh2011}. However, even a purely collision-less DM halo may relax very effectively via the
so-called violent relaxation \citep{violent}. The essence of this mechanism is simple: when the
halo collapses, strong density inhomogeneities (caustics etc.) should appear. The inhomogeneities
generate a small-scale gravitational field, that is a mediator of the energy exchange between
particles. Analytical calculations show that the mechanism can be very effective in the center of
the halo. However, the violent relaxation 'works' only during the halo collapse: the formed halo
has a stationary gravitational field. Moreover, the efficiency of the violent relaxation rapidly
drops with radius \citep{violent}.

Here we investigate an another, quite unphysical, relaxation mechanism: the pair collisions of the
test bodies. In the case of real systems this process is completely absent: the particle mass of
real dark matter is so small that their gravitational collisions play no role, as we will see
below. On the contrary, the test bodies are quite massive ($\sim 10^{-6}$ of the total halo mass)
and may effectively collide redistributing their energy and momentum. In reality, where the dark
matter particles carry $< 10^{-60}$ halo masses, this process is irrelevant. So, do N-body
collisions contribute to cusp formation in N-body simulations?

An exact investigation of the collision influence is a very complex task: it requires careful
consideration of the particle velocity and spatial distributions. However, a simple (but quite
reliable) and commonly-used method to estimate the collisional relaxation time is to use
'characteristic', averaged values of the particle velocities $v$ and radii $r$, instead of real
distributions \citep[eqn. 1.32]{bt}. Substituting the orbital time $r/v$ into this equation, we
obtain:
\begin{equation}
\dfrac{\langle\Delta v\rangle}{\delta t}\simeq 0\qquad \dfrac{\langle\Delta v^2\rangle}{\delta
t}\simeq \dfrac{8 v^2 \ln\Lambda}{N(r)}\cdot\dfrac{v}{r}
 \label{13a2}
\end{equation}
Here $N(r)$ is the number of particles inside radius $r$, $\ln\Lambda$ is the Coulomb logarithm.
Generally speaking, $\ln\Lambda$ depends on radius. According to \citet[chapter 1.2.1]{bt},
$\ln\Lambda=b_{max}/b_{min}$, where $b_{max}$ and $b_{min}$ are the characteristic maximum and the
minimum values of the impact parameter. $b_{max}$ is the maximum radius where the surface density
$N(r)/\pi r^2$ of the test bodies can be approximately considered as constant. $b_{min}$ is defined
by the radius where either the assumption of a strait-line trajectory breaks, or the newtonian
potential is no longer valid. Since $\ln\Lambda$ depends on $b_{max}$ and $b_{min}$ only
logarithmically, it is usually enough to make rough estimations of these quantities. For a stellar
system \citet[chapter 1.2.1]{bt} estimated $b_{max}\simeq \rvir$ and $b_{min}\simeq b_{90}$, where
$b_{90}$ is the $90$ degree deflection impact parameter. Then $\Lambda=\rvir/b_{90}\simeq \rvir
v^2/(Gm)\simeq N$, and this is the estimation of $\Lambda$ applied by \citet{power2003}.

In the case of N-body simulations, $b_{90}$ should not be smaller than the smoothing radius
$\varrho$. Second, the surface density rapidly drops at the halo edge. Let us consider the center
of NFW profile (\ref{13a1}), where $\rho\propto r^{-1}$: then the surface density $N(r)/\pi r^2$
remains constant up to $r=r_s$, and $b_{max}\simeq r_s$. So $\Lambda\sim r_s/\varrho$ in this
important instance; \citet{klypin2013} obtained $\Lambda= 3 r_s/\varrho$, i.e. $\Lambda$ in the
center of an NFW halo depends almost not at all on radius. The dependence may be more significant
for other profiles; however, $\ln\Lambda$ cannot be a strong function of radius in the central
region: it depends on $b_{max}$ and $b_{min}$ only logarithmically. Moreover, the gravitational
friction force acting on a particle, which is proportional to $\ln\Lambda$, is apparently neither
zero nor infinite in the halo center. Consequently, $\ln\Lambda$   is finite, but not equal to $0$
at $r=0$. For reasons of symmetry, $\ln\Lambda$ has an extremum at $r=0$ and so cannot cannot be a
sharp function near this point in the very general case. In the important instance of NFW profile
$\ln\Lambda$ is quite constant, as we could see. For a halo of concentration $\cvir=10$ and a
typical value $\varrho\simeq 10^{-3} \rvir$ we obtain $\ln\Lambda\simeq 6$.

The relaxation time $\tau_r$ is roughly defined by the moment when $\Delta v^2\simeq v^2$, and we
obtain from (\ref{13a2})
\begin{equation}
\tau_r= \dfrac{N(r)}{8 \ln\Lambda}\cdot\dfrac{r}{v}
 \label{13a3}
\end{equation}
Since real DM haloes contain $\sim 10^{65}$ particles, the collisional relaxation plays no role
there. On the contrary, the number of test bodies in a separate halo is relatively small in
simulations, and the unphysical relaxation may be important on the simulation time $t_0$.

\subsection{Convergence}
The main purpose of the convergence test is to find the maximum ratio $t_0/\tau_r$ whereby the
density profile is still not corrupted by the collisions. $\tau_r$ rapidly grows with radius: the
influence of the collision relaxation may still be negligible on the halo outskirts, but already
significant in the center. So we may introduce the convergence radius $r_{conv}$ of a halo at given
$t_0$, so that the collisions are already significant inside $r_{conv}$, but the simulated density
profile is still reliable for $r>r_{conv}$.

Of course, this aspect of the N-body simulation convergence has been explicitly explored; we
mention here only a few works of the vast literature. Because of the lack of reliable analytical
predictions of dark matter distribution near the halo center, the main method of the tests is to
find the conditions under which the structure of simulated halos is independent of numerical
parameters. \citet{moore1998} found that $r_{conv}\sim R_{vir}/\sqrt[3]{N_{vir}}$ ($R_{vir}$ and
$N_{vir}$ are the virial radius of the halo and the number of the test particles in it), and so
$r_{conv}$ should contain thousands test particles. \citet{ghigna2000} gave a similar estimation of
the particle number, but emphasized that $r_{conv}$ should exceed the smoothing radius $\varrho$ by
no less than a factor of three. On the contrary, \citet{klypin2001} suggested that it could be
enough to have $\sim 200$ simulative particles inside $r_{conv}$ if we properly chose other
simulation parameters. \citet{klypin2013} considered the density profiles of subhaloes and found
that even $\sim 100$ test particles can be enough in this case, because of the relatively small
number of crossing-times of dark matter in the subhaloes. \citet{diemand2004} argued for the
opposite: the two-body collisions in the subhaloes can be so effective that it affects the whole
hierarchical structure formation.

However, \citet{power2003} remains the fundamental investigation of the N-body convergence; the
criterion offered by this paper is routinely used in modern simulations to determine $r_{conv}$
(see, for instance, \citep{navarro2010}). In order to determine $r_{conv}$, \citet{power2003}
considered the time dependence of the overdensity at some radius $r$ (i.e., the ratio of the
average density inside radius $r$ from the halo center to the average universe density). It was
found that a cuspy profile (close to $\rho\propto r^{-1}$) forms fairly rapidly in the center, and
then the density contrast remains almost constant up to, at least, $t_0\simeq 1.7 \tau_r$
(\citet{power2003} used relaxation time, based on (\ref{13a3})). So the criterion of $r_{conv}$
offered by \citet{power2003} is $\tau_r\ge 0.6 t_0$. Such a large ratio ($t_0/\tau_r\simeq 2$)
seems surprisingly high: actually, it means that test particle collisions have little effect on the
simulations even at a time interval exceeding the relaxation time. However, it was found in further
convergence tests that the criterion might be even softer: the NFW-like profile in the center
survives almost unchanged many times longer than the analytical estimations of $\tau_r$
\citep{hayashi2003, klypin2013}. If one accepts stability of the profile as a proof of convergence,
it leads to a very optimistic appraisal of the central resolution of N-body simulations.

The aim of this paper is to show a potential pitfall of the above-mentioned numerical methods of
the convergence tests. The very fact of the simulation results converging to some 'stationary' (or
quasi-stationary) solution does not guarantee that the results have converged to a physical
solution. Indeed, we will show that test body collisions may result in the persistent occurrence of
the steep NFW-like central profile in numerous N-body simulations. If this is confirmed to be true,
the above-mentioned criteria of convergence (like $\tau_r\ge 0.6 t_0$), routinely used in
literature, would significantly overestimate the central resolution of N-body simulations.

\section{Calculations}
\subsection{The momentum distribution of the test bodies}

Let us consider test body collisions in the center of formed halo. We suppose a power-law density
distribution in the center $\rho\sim r^{-\beta}$ (it is important that $\beta<2$). Then the number
of the bodies enclosed inside radius $r$ and the halo mass depend on the radius as:
\begin{equation}
N(r)=N_0 \left(\frac{r}{r_b}\right)^{3-\beta}\qquad M(r)=\mu N_0
\left(\frac{r}{r_b}\right)^{3-\beta}
 \label{13a4}
\end{equation}
where $\mu$ is the test body mass, $r_b$ is the radius where the central profile $\rho\sim
r^{-\beta}$ breaks; $r_b=r_s$ in the case of NFW profile. The gravitational potential is
\begin{equation}
\phi(r)=G\dfrac{\mu N_0}{(2-\beta)r_b}
\left(\frac{r}{r_b}\right)^{2-\beta}=\dfrac{\Phi^2}{(2-\beta)}\left(\frac{r}{r_b}\right)^{2-\beta}
 \label{13a5}
\end{equation}
Assuming $\phi(0)=0$ and designating $\Phi=\sqrt{\dfrac{G \mu N_0}{r_b}}$. We follow the
simplifying method applied by \citet{bt} to obtain equation (\ref{13a2}): instead of exact
distribution functions, we will use characteristic, averaged over time values of particle radius
$r$, velocity $v$ and momentum $p=\mu v$. In the framework of this approach the task becomes
one-dimensional: the system state is totaly defined by a single function $N(r)$ depending on the
only coordinate ($r$, for instance). Of course, the halo center is virialised. The average kinetic
energy of a particle is $\ep_k=p^2/(2\mu)$, the average potential one $\ep_p=\mu \phi(r)$.
Potential (\ref{13a5}) is a power-law of index $(2-\beta)$: according to the virial theorem (see
\ref{13appendix1}), $\ep_k$ and $\ep_p$ of a particle being in finite motion in this field are
bound
\begin{equation}
\ep_k=\dfrac{2-\beta}{2}\ep_p
 \label{13b1}
\end{equation}
So
\begin{equation}
\dfrac{p^2}{2\mu}=\dfrac{2-\beta}{2}\dfrac{\mu
\Phi^2}{(2-\beta)}\left(\frac{r}{r_b}\right)^{2-\beta}\nonumber
\end{equation}
\begin{equation}
p=\mu \Phi \left(\frac{r}{r_b}\right)^{1-\frac{\beta}{2}}\qquad r=r_b \left(\frac{p}{\mu
\Phi}\right)^{\frac{2}{2-\beta}}
 \label{13a6}
\end{equation}
The last equation defines the one-to-one relation between the characteristic values of $r$ and $p$,
and we may use $p$ instead of $r$. In particular, $v=p/\mu$. Considering all particles contained
within radius $r$, their momentum distribution can be calculated as
\begin{eqnarray}
 \label{13a7}
 N(p)=N_0 \left(\frac{p}{\mu \Phi}\right)^{\frac{6-2\beta}{2-\beta}}\\
  \label{13a8}
 n(p)\equiv
\frac{dN(p)}{dp}=\frac{(6-2\beta)}{(2-\beta)}\frac{N_0}{\mu \Phi} \left(\frac{p}{\mu
\Phi}\right)^{\frac{4-\beta}{2-\beta}}
\end{eqnarray}
\subsection{The effect of collisions on the momentum distribution}
The collisions lead to a sort of diffusion of the test bodies in the phase space. Since the close
encounters are excluded by the potential smoothing, the evolution of distribution function $n(p)$
can be described by the Fokker-Planck equation:
\begin{equation}
\frac{\partial n(p)}{\partial t}=\frac{\partial }{\partial p}\left\{ {\tilde A}
n(p)+\frac{\partial}{\partial p}[B n(p)]\right\}
 \label{13a9}
\end{equation}
where
\begin{equation}
{\tilde A}=\dfrac{\langle\Delta p\rangle}{\delta t}= \mu\dfrac{\langle\Delta v\rangle}{\delta
t}\qquad B=\dfrac{\langle\Delta p^2\rangle}{2 \delta t}=\frac{\mu^2}{2}\dfrac{\langle\Delta
v^2\rangle}{\delta t}
 \label{13a10}
\end{equation}
Substituting here equations (\ref{13a2}), we obtain $\tilde A=0$ and
\begin{equation}
 B=\dfrac{4 p^3 \ln\Lambda}{r \mu N}=\dfrac{4 \mu^2 \Phi^3 \ln\Lambda}{r_b N_0} \left(\frac{p}{\mu
\Phi}\right)^{-\frac{2+\beta}{2-\beta}}
 \label{13a11}
\end{equation}
The diffusion flux of the particles is \citep{ll10}
\begin{equation}
s=-{\tilde A} n(p)-\frac{\partial}{\partial p}[B n(p)]
 \label{13a12}
\end{equation}
Since our task is one-dimensional, the number of particles crossing the surface $p=\text{\it
const}$ (or, what is the same in view of one-to-one relation (\ref{13a6}), $r=\text{\it const}$) in
a unit time is $dN(r)/dt=-s$. Substituting eqn.~(\ref{13a11}) for $B$ and eqn.~(\ref{13a8}) for
$n(p)$ into (\ref{13a12}), we obtain
\begin{equation}
-s=\frac{16(3-\beta)(1-\beta)}{(2-\beta)^2}\frac{\Phi\ln\Lambda}{r_b}\left(\frac{p}{\mu
\Phi}\right)^{-\frac{\beta}{2-\beta}}
 \label{13a13}
\end{equation}
In order to estimate the timescale $\tau_{col}$ influence of collisional diffusion on the density
profile, we should consider the ratio of the number of particles diffused through radius $r$ in
simulation time $t_0$ to the particle number $N(r)$ inside this radius, i.e. $\tau_{col}=N(r)\left/
\left|\frac{dN(r)}{dt}\right|\right.$. Substituting here $\tau_r(r)$ instead of $N(r)$ in
accordance with (\ref{13a3}), we obtain from (\ref{13a13})
\begin{equation}
\tau_{col}=\frac{(2-\beta)^2}{2(3-\beta)|1-\beta|} \tau_r
 \label{13a15}
\end{equation}
It is remarkable that the typical evolution time in units of $\tau_r$ depends only on $\beta$, and
not on radius.

\begin{figure}
 \resizebox{\hsize}{!}{\includegraphics[angle=0]{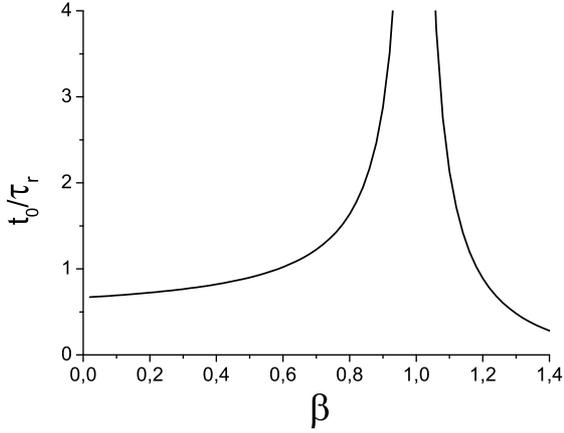}}
\caption{The time $t_0$ the central power-law profile $\rho\propto r^{-\beta}$ takes to be
significantly changed by Fokker-Planck diffusion (in units of relaxation time $\tau_r$, see
eqn.~(\ref{13a15})). As we can see, if $\beta$ significantly differs from $1$, the profile evolves
towards $\rho\propto r^{-1}$ in a time shorter than $\tau_r$. On the contrary, profile $\rho\propto
r^{-1}$ survives for a much longer time interval ($t_0\gg\tau_r$).}
 \label{13fig1}
\end{figure}

Obviously, $\tau_{col}\gg \tau_r$ for $\beta\simeq 1$, implying that the system evolves very slowly
on account of collisions, despite being collisional, if the profile is approximately $\rho\propto
r^{-1}$. This result suggests that stability of a $r^{-1}$ profile does not guarantee the absence
of collisional effects. In fact, collisions might provide a stable $r^{-1}$ density profile.
Moreover, profile $\rho\propto r^{-1}$ is a sort of attractor: we may see from (\ref{13a13}) that
the diffusion flux changes its sign at $\beta=1$. If the profile is shallower than $\rho\propto
r^{-1}$, the flux is pointed towards the center and tends to sharpen it, and vice versa, the flux
shallows a profile with $\beta>1$. Fig.~\ref{13fig1} illustrates the characteristic time necessary
for the central part of the system to converge towards the $\rho\propto r^{-1}$ profile. As we can
see, if $\beta$ significantly differs from $1$, the profile evolves in a time shorter than
$\tau_r$. On the contrary, profile $\rho\propto r^{-1}$ may survive for a much longer time interval
($t_0\gg\tau_r$).

\section{Discussion}
Thus we have come to the main conclusion of this paper: the very fact of the simulation results
converging to some 'stationary' (or quasi-stationary) solution does not guarantee that the results
have converged to the physical distribution. Numerical effects may also create some stable
pseudosolutions.

\subsection{The immanence of a NFW-like attractor solution}

Of course, our model was rather crude: we considered $\beta$ as a constant: so we can only
ascertain the characteristic time, in which the profile will significantly change, and the
direction of evolution, being unable to follow the evolution in time. Moreover, we used
'characteristic', averaged values of the particle velocities $v$ and radii $r$, instead of real
distributions; our analysis certainly fails if $r\lesssim\varrho$. This simplified approach allowed
us to follow exactly the collision estimation method utilized by \citet{power2003} and to make the
calculations short and obvious.

However, the presence of a NFW-like attractor solution ($\rho\propto r^{-\beta}$, where
$\beta\simeq 1$) in the halo center seems to be an immanent result of the particle collisions,
independent on the simplifications we made. Indeed, \citep{spitzer1987, quinlan1996} consider a
spherical star cluster with the distribution function depending only on energy and time (so the
velocity distribution is assumed to be isotropic). The diffusion coefficients of the Fokker-Planck
equation are evaluated in the local approximation. Though the equations are much more complex in
this case, an attractor solution, very similar to that obtained in this paper, occurs as well
\citep{evans1997}. The only difference is that the power index is $\beta=4/3$ instead of NFW's
value $\beta=1$. However, the discrepancy does not seem very significant. First, the convergency
radius $r_{conv}$ typically contains hundreds or thousands of test particles in the case of N-body
simulations, and so the statistical noise is rather high and does not allow to reconstruct the
density profile and determine $\beta$ at $r\sim r_{conv}$ very precisely. Second, the model
\citep{spitzer1987, quinlan1996} is also not precise: it calculates the diffusion coefficients in
the local approximation (which is hardly suitable for an N-body halo at $r\sim r_{conv}$ with only
a moderate number of particles inside) and disregards the potential smoothing. Finally, there is an
effect that shallows the attractor profile, making it even closer to NFW.

\subsection{The cusp disappearance at $t\sim 50\tau_r$}
As was emphasized in \citep{evans1997}, condition $s=0$ (i.e., the cessation of the Fokker-Planck
diffusion) means the dynamical, but not thermodynamical equilibrium of the system. The full
Boltzmann equation has the only stationary solution --- the isothermal sphere. The Fokker-Planck
equation is obtained by expanding the full Boltzmann collision integral into Taylor's series and
dropping out all the terms beyond second order \citep{ll10}. Therefore a stationary solution of the
Fokker-Planck equation is not obligatory stationary with respect to the full collision integral.
Indeed, the particle velocity dispersion (and so the 'temperature') drops towards the halo center
for any density profile softer than $\rho\propto r^{-2}$ (see, for instance, the first part of
equation (\ref{13a6})). So the NFW-like solutions, like those considered in this paper or in
\citep{evans1997}, have a nonuniform temperature distribution, and the particle collisions should
still redistribute energy, leading to additional transport processes. However, the fact that we
consider a stationary solution of the Fokker-Planck equation results in an important consequence:
the main terms of the collision integral are self-compensated in this case. So the transport
processes are possible, but they are caused by the higher-order terms of the collision integral,
which are notably smaller. The attractor solutions should survive for much longer than any other
central distribution, i.e. for much longer than $\tau_r$. However, they are finally smeared out by
the kinetic terms of higher orders. Since the process is caused by the collisions and by the
nonuniform distribution of the system 'temperature', this effect may be considered as an
'additional thermal conductivity'.

We can estimate the lifetime of the NFW-like attractor profiles. According to \citep[eqn.
22]{evans1997}, the heat flux $Q=dE/dt$ towards the center in the case of the attractor solution is
\begin{equation}
Q=\dfrac{25\pi\log\Lambda}{768\sqrt2}\dfrac{\mu \phi^{3/2}(r)}{r}
 \label{13a16}
\end{equation}
In the case under consideration $\beta=4/3$. With the help of (\ref{13b1}), we can estimate
$\phi(r)\simeq \frac{2}{2-\beta} \frac{v^2}{2}=\frac32 v^2$ and the total kinetic energy of the
particles inside radius $r$: $E_{tot}(r)\simeq N(r) \frac{\mu v^2}{2}$. Substituting these
estimations and equation (\ref{13a3}) for $\tau_r$ into (\ref{13a16}), we obtain:
\begin{equation}
Q=\dfrac{25\pi\sqrt3}{8192}\dfrac{E_{tot}(r)}{\tau_r}
 \label{13a17}
\end{equation}
The ratio $\tau_{th}=E_{tot}/Q$ gives the time scale $\tau_{th}$ on which the 'additional heat
flux' significantly affects the system. The 'heat' relaxation time is much larger than the
collisional one, $\tau_{th}\simeq 60 \tau_r$, and this ratio does not depend on the radius. As we
can see, the attractor solutions survive much longer then $\tau_r$. However, they are eventually
smeared out by the higher order terms of the kinetic equation at tens of relaxation times.

Indeed, \citet{hayashi2003} and \citet{klypin2013} report that the NFW-like cusp smears out, when
$t_0$ reaches several tens of $\tau_r$. They considered this effect as the first sign of the
influence of the test body collisions. However, our calculations suggest that collisions become
important much earlier, at $t_0\lesssim\tau_r$: it may, in fact, be the collisions that form the
cusp itself. The smearing of the cusp at $t_0\sim 50$ may be a result of the 'additional thermal
conductivity': the characteristic time and the sense of the process coincide. The additional heat
flux, being also a result of the test body collisions, is much slower than the Fokker-Planck
diffusion and becomes important approximately $50$ times later.

Thus the property that the particle encounters form a cuspy, NFW-like profile in the center of the
halo in a time interval $\sim \tau_r$ seems to be quite general and model-independent. Since this
is an attractor solution, the profile will survive for $t_0\gg \tau_r$. However, considering that
the process is driven by the unphysical test-body collisions, it would be an artefact of the
simulations. The NFW profile may still be valid. If the halo concentration is small ($c_{vir}\sim
3$), $\rho\propto r^{-1}$ profile occurs at large radii, where certainly $t_0\ll\tau_r$, and the
collisional relaxation has nothing to do with it. Perhaps, the violent relaxation (as we could see,
a strong energy relaxation is mandatory to form the cusp) may form the same profile even closer to
the center, where $t_0\ge\tau_r$. However, the simulation results require further verification if
$t_0\simeq\tau_r$: the Fokker-Planck diffusion may produce the same profile as well. Since there is
no obvious way to distinguish these two processes, the criterion $t_0\ll\tau_r$ seems to be the
only reliable condition of the negligibility of the unphysical numerical effects, while criterion
$\tau_r\ge 0.6 t_0$ \citep{power2003} significantly overestimates the resolution of N-body
simulations in the halo center.

\subsection{How to test the veracity of the
simulation results?}

Our suggestions could eliminate the 'cusp vs.\ core' problem, which actually appears mainly on the
scale of galaxies and dwarf satellites. First, N-body simulations results come into conflict with
the observational data only if soft criteria like $\tau_r\ge 0.6 t_0$ are true. As we saw, we
probably should use a more conservative criterion $t_0\ll\tau_r$, which increases several times the
convergence radius. Then the numerical resolution can be just too low to resolve cores in the
centers of the galaxy haloes, that are typically a few percent of the virial radius.

Second, a supposition of the moderate energy relaxation of galactic haloes inevitably leads to the
central profile that fits observational data much better than the cuspy profiles suggested by
N-body simulations: the profile has a core, an extensive region with $\rho \propto r^{-2}$, and the
product of the central density and the core radius is almost independent of the halo mass
\citep{16}. On the contrary, N-body simulations suggest a very strong energy relaxation of the
system \citep{moore1999, mo09}. However, we could see that the residual test body collisions might
boost the energy exchange in the halo center, leading to disagreement with the observations.

Thus the criteria based on the stability of the central profile, like \citep{power2003}, seem to be
questionable. The absence of the collisional relaxation is guaranteed only if $t_0\ll\tau_r$. A
question appears: how can we quantitatively estimate the influence of the collisions and optimize
the limit for $t_0/\tau_r$? The most direct way is to consider the evolution of the total particle
energy $\ep$ depending on their average radius and time. Indeed, violent relaxation is effective
only during the short interval of halo collapse \citep{violent}, after which the gravitational
field of the halo becomes stationary, and the total energy of each particle remains more or less
constant. There are several factors that can yet affect the particle energy distribution of the
formed halo.
\begin{enumerate}
 \item\label{13item1} An already formed halo may gravitationally capture more substance from the surroundings
 \citep{wang2012, 14}. The accretion of the additional mass can lead to the 'secondary violent relaxation'
 \item Tidal influence of the nearby halos
 \item \label{13item2} The substructures
 \item \label{13item3} A slow evolution of the formed halo, like its concentration growth
 \item \label{13item4} Unphysical test body collisions
\end{enumerate}
The first four items correspond to quite physical phenomena, while the fifth one is purely
numerical and should be avoided. As we could see, the collisions appear to be the most important in
the halo center, so we should focus our attention on this area. The central region contains few if
any subhaloes, as they are typically tidally destroyed. The tidal influence of the nearby halos is
minor because of the smallness of the region comparing to the halo size. We can completely avoid
the influence of late accretion (item \ref{13item1}) and the tidal influences by considering an
isolated halo. Moreover, even in the case of more sophisticated simulations containing many haloes,
the fraction of matter accreted to the central area after the main relaxation and the presence of
subhalos are easily measurable. Thus we may estimate the influence of factors
\ref{13item1}-\ref{13item2}.

In the end, we may wish to discriminate between the physical process \ref{13item3} and the
unphysical process \ref{13item4}. In the following we propose an independent test of the absence or
presence of significant test-body collisions in numerical simulations, that can be performed in
real time as the simulations progress. First of all, we should determine the total energy $\ep$ of
each particle at the moment, when the halo has just been formed, and its gravitational field
becomes stationary. A marker of this moment could be the attenuation of initial density waves and
caustics, i.e. the absence of rapid oscillations of the density profile and the gravitational
potential at the halo center. Then we can extract from the particle distribution over the
fractional variation $\Delta\ep/\ep$ of their total energies as a function of time and the average
radius of particles. The effects of processes \ref{13item3} and \ref{13item4} on the energy
evolution differ drastically. Process \ref{13item3} leads to a slow regular shift of the particle
energies. The energies of the bodies with similar orbits evolve alike. On the contrary, process
\ref{13item4} leads to stochastic variations of the particle energy. The relative energy change
$\Delta\ep/\ep$ should have a near-Gaussian distribution with dispersion $\sigma$ roughly
proportional to $\sqrt{t}$, i.e. $\sigma\simeq \sqrt{t/\tau_r}\propto \sqrt{t/N(r)}$. The energy
evolution driven by the unphysical test body collisions is stochastic and can be distinguished from
the physical processes like (\ref{13item1}-\ref{13item3}). A more sophisticated method to
investigate the influence of the collisions is to consider the adiabatic invariants associated with
the particles. Indeed, the quantities $\oint p_i dq_i$ (where $q_i$ and $p_i$ are a generalized
coordinate and the corresponding generalized momentum of the particle) are conserved, if the
gravitational field changes slowly \citep{ll1}. On the contrary, particle collisions should lead to
accidental variations of the adiabatic invariants.

One can offer several practical ways to test the influence of the unphysical particle collisions in
real N-body simulations. Any method has its strong and weak points. We can consider standard
cosmological simulations containing many halos and choose a halo of interest. Then we need to find
a moment $t_{in}$ when the halo density distribution becomes almost stationary. Even the most
massive halos are quite formed at $z=1$ in the $\Lambda$CDM cosmological model. We remind that
$z=1$ corresponds to $\sim 40\%$ of the present Universe age, that is, a halo has a
quasi-stationary density profile and gravitation field during the major part of its physical age.
Then we need to consider the energy evolution of the test bodies
$|\ep(t)-\ep(t_{in})/\ep(t_{in})|$, as this was described above. A slow physical evolution of the
halo, such as the concentration growth, leads to a regular energy change similar for all the
particles that can be easily estimated by the gravitational potential change. A random walk of the
particle energies undoubtedly reveals the influence of the unphysical collisions. A possible
difficulty of this method is the fact that halos in cosmological simulations are never spherically
symmetric. This significantly complicates the task making it three-dimensional. In particular, one
needs to perform accurate calculations of the gravitational potential going beyond the
commonly-used spherical approximation.

The second way to estimate the influence of the collisions is a simulating of of one of the
well-known analytical models of s stationary isolated halo, such as Plummer, Hernquist, or Osipkov
distributions \citep{bt}. The particle velocity distribution and the density profile are
time-independent in this case. There should be no relaxation, since the gravitational field is
stationary. Therefore, any evolution of the distribution function (in particular, the cusp
formation) would be a clear sign of numerical effects.

Both above-mentioned testing methods are only applicable for a formed halo and unsuitable for the
halo collapse consideration. However, test body collisions may cause undesirable numerical effects
during the collapse as well. Indeed, in the well-known case of Tolman collapse (the initial
perturbation is spherically symmetric and uniform $\rho=\text{\it const}$, the matter has no
angular momentum \citep{ll2}) all the particles reach the center simultaneously, and the halo
density becomes infinite. In the instance of real initial perturbation, $\rho$ grows towards the
center and particle angular momentum is not exactly zero. However, in the very general case the
initial density contrast is low $\delta\rho/\rho\ll 1$. The initial angular momentum is zero
\citep{gorbrub2}, though it can be gained later by the tidal interaction. Therefore, the collapse
dynamics is similar in the early phase to the Tolman case, that is, any real halo passes through a
stage of high compression at the beginning of the collapse. Though the stage is fairly short, the
test particle collisions may significantly redistribute the particle energy, since the collision
rate is proportional to $\rho^2$. Thus the collisions of test bodies make an unphysical
contribution to the violent relaxation in simulations. Moreover, a halo during the violent
relaxation has a complex structure of the density and gravitational field. The ability of
simulations to model adequately the fine structure of numerous caustics and other inhomogeneities
by a fairly small number of test bodies is questionable, while the efficiency of relaxation
strongly depends on this. Thus the problem of the reliability limits of the N-body simulations
needs further consideration. The profile stability criterion along is insufficient.

The convergence criteria obtained with the help of a direct and detailed consideration of the
energy evolution of the system should be significantly more reliable than those based on the
density profile stability. The stationary solutions obtained in this paper or by \citet{evans1997}
serve as a good illustration. The attractor density profiles are quite stable even if $t\gg\tau_r$,
though the unphysical collisions of the test bodies form them. Despite of the profile stability,
the energies of the particles exhibit accidental variations, revealing the collision influence.
Equations (\ref{13a2}), (\ref{13a3})) show that $\Delta\ep/\ep\simeq 1$ when $t=\tau_r$, i.e. the
random energy fluctuations are fully visible. Thus the insensitivity of the density profile to the
simulation parameters is not a sufficient criterion of convergence per se, and a more detailed
consideration of the phase evolution of test bodies is necessary to judge the simulation
reliability. We have proposed a simple test that can be performed in any simulation code.

\section{Acknowledgements}
 Financial support by Bundesministerium f\"ur Bildung und Forschung through DESY-PT, grant
05A11IPA, is gratefully acknowledged. BMBF assumes no responsibility for the contents of this
publication. We acknowledge support by the Helmholtz Alliance for Astroparticle Physics HAP funded
by the Initiative and Networking Fund of the Helmholtz Association.

\appendix
\section{The virial theorem for the case of a single particle moving in a potential field $\phi(r)$}
\label{13appendix1} Let us denote by $x_i$ and $p_i$ the three components of the particle radius
and momentum. Then (we use the Einstein summation convention: when an index variable appears twice
in a single term it implies summation of that term over all the values of the index.)
\begin{equation}
\frac{d (p_i x_i)}{dt}= x_i\frac{d p_i}{dt}+ p_i\frac{d x_i}{dt}= -m x_i \frac{d \phi}{dx_i}+ 2 T
 \label{13app1}
\end{equation}
Here we took into account that $\frac{d p_i}{dt}=F_i=-m \frac{d \phi}{dx_i}$ and $p_i\frac{d
x_i}{dt}=p_i v_i=2T$, where $T$ is the particle kinetic energy. We may introduce the time averaging
of a quantity $A$: $\bar A\equiv \frac{1}{t}\int^t_0 A dt$ and apply it to eqn. \ref{13app1}.
\begin{equation}
\frac{(p_i x_i)_t - (p_i x_i)_0}{t}=  -m \overline{\left(x_i \frac{d \phi}{dx_i}\right)}+ 2
\overline{T}
 \label{13app2}
\end{equation}
If the particle motion is finite, $(p_i x_i)$ is also finite, while $t$ can be arbitrarily large.
So, the left part of the equation approaches zero for $t\to\infty$. Eqn. \ref{13app2} may be
significantly simplified, if $\phi$ is a homogeneous function, i.e., $\phi(\lambda x_1, \lambda
x_2, \lambda x_3)=\lambda^a \phi(x_1, x_2, x_3)$. Then $x_i \frac{d \phi}{dx_i}=a \phi$. Since
$m\phi$ is the potential energy of the particle $\Pi$, we obtain
\begin{equation}
 \frac{a}{2}\; \overline{\Pi}=\overline{T}
 \label{13app3}
\end{equation}
In particular, if $\phi(r)\propto r^{(2-\beta)}$, the potential is a homogeneous function with
$a=2-\beta$.


\end{document}